\documentclass[prl,twocolumn,superscriptaddress,nofootinbib,aps,10pt,preprintnumbers,
longbibliography]{revtex4-1}

\usepackage[colorlinks=true,linkcolor=black,citecolor=blue,urlcolor=blue, pdfborder={0 0 0}]{hyperref}
\usepackage{mathtools}
\usepackage[utf8]{inputenc}
\usepackage{color}
\usepackage[table,xcdraw,dvipsnames]{xcolor}
\usepackage{multirow}
\usepackage{slashed}
\usepackage{mathbbol}
\usepackage[normalem]{ulem}    

\usepackage{tikz}
\usetikzlibrary{shapes,arrows,positioning,automata,backgrounds,calc,er,patterns}
\usepackage[compat=1.1.0]{tikz-feynman}

\newcommand{\beq} {\begin{equation}}
\newcommand{\eeq} {\end{equation}}
\newcommand{\bea} {\begin{eqnarray}}
\newcommand{\eea} {\end{eqnarray}}
\newcommand{\ba} {\begin{eqnarray*}}
\newcommand{\ea} {\end{eqnarray*}}

\definecolor{Celadon}{rgb}{0.67, 0.94, 0.82}
\definecolor{Pink}{rgb}{0.9, 0.0, 0.0}
\definecolor{darkred}{rgb}{0.5, 0.2, 0.13}
\definecolor{ForestGreen}{RGB}{34,139,34}

\newcommand{\titleprl}[1]{\vspace{0.3cm} \noindent \textit{\textbf{#1}}:}

\begin{document}

\preprint{TTP21-009, P3H-21-025}
\title{Testable dark matter solution within the seesaw mechanism}

\newcommand{\affME}{{\small \it  Dipartimento di Scienze Matematiche e Informatiche, Scienze Fisiche e Scienze della Terra, \\ Universit\`a degli Studi di Messina, Viale Ferdinando Stagno d'Alcontres 31, I-98166 Messina, Italy}}
\newcommand{\affCAT}{{\small \it INFN Sezione di Catania, Via Santa Sofia 64, I-95123 Catania, Italy}}
\newcommand{\affBO}{{\small \it Dipartimento di Fisica e Astronomia, Universit\`a di Bologna, via Irnerio 46, 40126 Bologna, Italy}}
\newcommand{\affINFNBO}{{\small \it INFN, Sezione di Bologna, viale Berti Pichat 6/2, 40127, Bologna, Italy}}
\newcommand{\affORSAY}{{\small \it P\^ole Th\'eorie, Laboratoire de Physique des 2 Infinis Irène Joliot Curie (UMR 9012), \\
CNRS/IN2P3 et Universt\'e Paris Saclay, 
15 Rue Georges Clemenceau, 91400 Orsay, France}}

\author{Asmaa Abada}
\email{asmaa.abada@ijclab.in2p3.fr}
\affiliation{\affORSAY}

\author{Giorgio Arcadi}
\email{giorgio.arcadi@unime.it}
\affiliation{\affME}
\affiliation{\affCAT}

\author{Michele Lucente}
\email{michele.lucente@unibo.it}
\altaffiliation[\\Also visitor at: ]{\emph{Theoretical Physics Department, Fermi National Accelerator Laboratory, Batavia, Illinois 60510, USA}}
\affiliation{\affBO}
\affiliation{\affINFNBO}

\author{Salvador Rosauro-Alcaraz}
\email{rosauroa@bo.infn.it}
\affiliation{\affINFNBO}

\begin{abstract}
\noindent 
The dark matter problem, together with the discovery of neutrino masses from the observation of the oscillation phenomenon, represents one of the most important open questions in particle physics. A concurrent solution arises when one of the right-handed neutrinos, necessary for the generation of light neutrino masses, is the dark matter candidate. In this letter, we study the generation of such a dark matter candidate relying solely on the presence of neutrino mixing. This tightly links the generation of dark matter with searches in laboratory experiments on top of the usual indirect dark matter probes. We find that the regions of parameter space producing the observed dark matter abundance can be probed with electroweak precision observables, charged lepton flavor violation searches, as well as, collider searches when the mass of the heavy neutrino lies at the TeV scale. 

\end{abstract}

\maketitle

\setcounter{footnote}{0}

\newenvironment{Appendix}
{
	\setcounter{section}{1}
	\setcounter{equation}{0}
	\renewcommand{\thesubsection}{\Alph{subsection}}
	\renewcommand{\theequation}{A.\arabic{equation}}
}


\titleprl{Introduction}
Despite its enormous success, 
the Standard Model (SM) of particle physics has a number of known shortcomings, such as the origin of neutrino masses, the lack of dark matter (DM) candidate in the Universe, and the generation of the baryon asymmetry (BAU), among others. A minimal extension of the SM simultaneously accommodating the aforementioned puzzles would be undoubtedly appealing. 


Fermionic seesaw-based extensions of the SM potentially have this feature, since under suitable conditions, right-handed (RH) neutrinos responsible for the generation of neutrino masses can be viable DM candidates. Furthermore, they can account for the BAU via leptogenesis~\cite{Fukugita:1986hr,Akhmedov:1998qx,Davidson:2008bu,Hernandez:2016kel,Abada:2017ieq,Sandner:2023tcg}. For instance, the 
$\nu$MSM~\cite{Asaka:2005an,Asaka:2005pn,Asaka:2006nq,Shaposhnikov:2008pf,Laine:2008pg} model can achieve the generation of neutrino masses and of the BAU, while providing a DM candidate produced resonantly in the presence of lepton asymmetries~\cite{Canetti:2012kh,Ghiglieri:2019kbw,Ghiglieri:2020ulj}.

We propose here a different DM production mechanism, while complying with neutrino mass generation when extending the SM with three RH neutrinos: the DM relic density is produced ``\`a la'' freeze-in through two body decays, without relying on any preexisting lepton asymmetry. Moreover, the non-DM RH neutrinos need to lie at scales of the order of $100\,\mathrm{GeV}-1\,\mathrm{TeV}$ and have sizable couplings with the SM. The effectiveness of this mechanism places the viable parameter space in reach for future collider probes, such as FCC, and experiments searching for charged lepton flavor violation (cLFV).


The letter is organized as follows: after the introduction of the mechanism generating light neutrino masses, we describe the DM production mechanism in the context of thermal field theory (TFT), and the computation of its abundance. We then present relevant up-to-date  experimental constraints. Finally, we discuss our main results as well as future experimental prospects, and conclude.

\titleprl{Theoretical setup}
We embed onto the SM the Type-I seesaw~\cite{Minkowski:1977sc,Mohapatra:1979ia,Yanagida:1979as,Gell-Mann:1979vob} with three RH 
neutrinos $N_{R_i}$, one of which\footnote{We choose the lightest of them, $N_{R_1}$, to be the DM candidate.}
represents the DM candidate with a mass $m_{\mathrm{DM}}\sim\mathcal{O}$(keV). The SM Lagrangian is completed by
\begin{equation}
    \mathcal{L}\supset -\bar{L}_L \tilde{\Phi} Y_{\nu} N_{R}-\frac{1}{2}\bar{N}_{R}^c M_N N_{R}+h.c.\,,
\end{equation}
$L_L$ ($\Phi$) being the $SU(2)_L$ lepton (Higgs) doublet, $Y_{\nu}$ is the neutrino Yukawa matrix and 
$M_N$ is the sterile neutrino Majorana mass matrix, assumed real and diagonal without loss of generality.
After electroweak (EW) spontaneous symmetry breaking (SSB), the light-neutrino masses, assuming $m_D\ll M_N$, are given by $m_{\nu}\simeq -m_D M_N^{-1}m_D^T$, where $m_D\equiv v_H Y_{\nu}/\sqrt{2}$ and $v_H$ is  the Higgs vacuum expectation value (vev).
We rewrite $m_D$ using  the Casas-Ibarra (CI) parameterization: 
\begin{equation}
    m_D=-i U_{\mathrm{PMNS}}\sqrt{m_d} R^T \sqrt{M_N}\,,
    \label{eq:Casas-Ibarra}
\end{equation}
where $U_{\mathrm{PMNS}}$ is the lepton mixing matrix measured in oscillation experiments, $m_d$ is the diagonal light-neutrino mass matrix and $R$  an orthogonal matrix, which  can be parameterized with three rotations $V_{ij}$ in the $i-j$ plane, $R=V_{23}V_{13}V_{12}$, and thus by three complex angles $\vartheta_{ij}$.

The CI parameterization guarantees the correct description of oscillation data~\cite{Esteban:2024eli} from the diagonalization of the full neutrino mass matrix.
We will work in the limit of approximate lepton-number conservation (LNC)~\cite{Branco:1988ex,Kersten:2007vk,Abada:2007ux,Moffat:2017feq,Lucente:2021har},\footnote{Scenarios relying on this symmetry argument to explain the smallness of light-neutrino masses are generally dubbed low-scale seesaw scenarios. Examples include the inverse~\cite{Schechter:1980gr,Gronau:1984ct,Malinsky:2005bi,Abada:2014vea,Abada:2014zra} and linear~\cite{Mohapatra:1986aw,Mohapatra:1986bd,Barr:2003nn} seesaw mechanisms, among others~\cite{Asaka:2005an,Asaka:2005pn,Shaposhnikov:2006nn}. They allow for large Yukawa couplings with $M_N$ around the EW scale.} which allows for large active-heavy mixing angles, $U_{\alpha (i+3)}$ with $\alpha\in \lbrace e,\,\mu,\,\tau\rbrace$ and $i\in\lbrace 1,\,2,\,3\rbrace$, while protecting light-neutrino masses from radiative corrections. 
$U_{\alpha (i+3)}$ controls the strength of the interactions between the heavy mostly-sterile states and the SM and is approximately given by $U_{\alpha (i+3)}\simeq (m_D M_N^{-1})_{\alpha i}$. In the CI parameterization, and depending on the light-neutrino mass ordering, the LNC limit 
is found for different values of $\vartheta_{12}$ and $\vartheta_{13}$ while having $\mathrm{Im}\left(\vartheta_{23}\right)\gg 1$. For normal ordering (NO), it corresponds to $\vartheta_{12},\,\vartheta_{13}\rightarrow 0$, while for inverted ordering (IO) it is realized when $\vartheta_{12},\,-\vartheta_{13}\rightarrow -\pi/2$. 

In the LNC 
limit one naturally finds that the DM candidate interacts very weakly, $|\theta_{\alpha \mathrm{DM}}|\equiv |U_{\alpha 4}|\ll 1$, in agreement with the strong bounds arising from X-ray searches~\cite{Boyarsky:2007ge,Roach:2019ctw,Foster:2021ngm,Roach:2022lgo}. Its contribution to light neutrino masses is very suppressed and thus the lightest neutrino is practically massless, while the two additional heavy states $N_{1,\,2}$, which are almost degenerate with a mass $m_{N}$, have large mixing angles satisfying $U_{\alpha N}\equiv U_{\alpha 5}\simeq i U_{\alpha 6}$.\footnote{In what follows, we label indices related to the heavy neutrinos that form a pseudo-Dirac pair with capital letters.} 
As will be shown, EW precision observables (EWPO) and searches for cLFV provide strong bounds on 
$|U_{\alpha N}|$ for $m_N$ larger than the Z-boson mass~\cite{Blennow:2023mqx}. For lighter masses, direct searches at colliders or beam dump experiments provide  the strongest constraints~\cite{ATLAS:2019kpx,ATLAS:2022atq,ATLAS:2024fcs,CMS:2018iaf,CMS:2022fut,CMS:2024ake,CMS:2024ita,CMS:2024xdq,Kelly:2021xbv,MicroBooNE:2023eef}.
\titleprl{Dark matter production rate}
We are interested in the DM production rate through two-body decays of the SM bosons into the DM, as well as decays of the heavy neutrinos 
into a SM boson and the DM. These can only take place after neutrino masses and mixings are generated.\footnote{There is a direct contribution to the DM production rate from Higgs decays even in the symmetric phase~\cite{Ghiglieri:2016xye}, but we neglect it in the following and focus on the production through mixing.} Indeed, after SSB we have interactions between the different neutrino mass eigenstates and the SM, as detailed in the supplemental material. Therefore, at temperatures $T\lesssim T_{\mathrm{SSB}}\simeq 160$~GeV, these decays involving the SM bosons can be considered.

The DM production rate can be obtained in a consistent way by computing the self-energy corrections to the neutrino propagator 
in the context of TFT~\cite{Schwinger:1960qe,Keldysh:1964ud,le_bellac_1996,kapusta_gale_2006,Lundberg:2020mwu}. 
Even if this idea has already been considered in the work of Ref.~\cite{Lello:2016rvl}, we have embedded it in a realistic scenario for neutrino mass generation based on the seesaw mechanism for the first time. This implies the inclusion of contributions not considered in Ref.~\cite{Lello:2016rvl}, notably those from the heavy neutrinos in the Higgs-mediated self-energy diagrams. We present 
for the first time 
a complete computation of such contribution, superseding the result of Ref.~\cite{Abada:2023mib}. In particular, we have determined analytically the DM production rate for each helicity ($h=\pm 1$) once the self-energy corrections have been computed. 


Our starting point is the Dirac equation for the neutrinos in a medium:
\begin{equation}
    \left(\slashed{p}-\mathcal{M}+\slashed{\Sigma}^L P_L+\slashed{\Sigma}^R P_R\right)\psi=0\,,
    \label{eq:Dirac_eq}
\end{equation}
where $\mathcal{M}$ 
is the diagonal neutrino mass matrix and
$\slashed{\Sigma}^{\chi}$ represents the self-energy correction associated to a given chirality projection $\chi=L,\,R$. Expanding the field $\psi$ in terms of helicity eigenstates~\cite{Lello:2016rvl,Abada:2023mib} and keeping the dominant self-energy contributions, we arrive at the following inverse propagator:\footnote{Details on this derivation are in the supplemental material.}
\begin{equation}
    \begin{split}
        \mathcal{S}_h^{-1}\equiv& p_0^2-p^2-\Omega_h=p_0^2-p^2+
        \left(p_0-h p\right)\Sigma^{L}\\
     &-\left(p_0-hp\right)\mathcal{M}
    \left[p_0-hp+\Sigma^{R}\right]^{-1}\mathcal{M}\,.
    \label{eq:Inverse_Propagator}
    \end{split}
\end{equation}
In Eq.~(\ref{eq:Inverse_Propagator}), 
$p_0$ and $p$ should be understood as matrices proportional to the identity in the neutrino mass basis, while  $\Omega_h$ corresponds to every term after $p_0^2-p^2$ in the second equality. Assuming the dispersion relation for the DM candidate can be approximated by $p_0^2-p^2-m_{\mathrm{DM}}^2-\delta\omega_h\simeq0$, with $|\delta\omega_h|/m_{\mathrm{DM}}^2\ll1$, the rate at which DM approaches equilibrium\footnote{This corresponds exactly to the DM production rate in Eq.~(\ref{eq:Boltzmann_Eq}).} corresponds to $\Gamma_{\mathrm{DM}}^h=-2\mathrm{Im}p_0\simeq -\mathrm{Im}\left[\delta\omega_h\right]/\sqrt{p^2+m_{\mathrm{DM}}^2}$. Neglecting light neutrino masses, we find the following DM equilibration rate for $h=+1$:
\begin{equation}
    \begin{split}
        \Gamma_{\mathrm{DM}}^{h=+1}\simeq& \frac{m_{\mathrm{DM}}^2}{2p^2}m_{\mathrm{DM}}^2\times\\
        &\mathrm{Im}\left[\mathcal{P}^R_{1N^{\prime}}(\mathcal{P}^R_{NN^{\prime}})^{-1}\mathcal{P}^R_{N1}-\mathcal{P}^R_{11}\right]\,,
    \end{split}
    \label{eq:DM_prodRate_hp1}
\end{equation}
while for negative helicity we arrive at
\begin{equation}
    \begin{split}
        \Gamma_{\mathrm{DM}}^{h=-1}\simeq &2\mathrm{Im}\big[\mathcal{P}^L_{11}\\
        -&2p\mathcal{P}^L_{1N^{\prime}}\left(m_N^2\mathbb{I}_{2}+2p\mathcal{P}^L_{NN^{\prime}}\right)^{-1}\mathcal{P}^L_{N1} \big]\,.
    \end{split}
    \label{eq:DM_prodRate_hm1}
\end{equation}
In Eq.~(\ref{eq:DM_prodRate_hm1}), $\mathbb{I}_2$ is the $2\times2$ identity matrix while $\mathcal{P}^R$ and $\mathcal{P}^L$ in Eqs.~(\ref{eq:DM_prodRate_hp1}-\ref{eq:DM_prodRate_hm1}) are $3\times3$ matrices that we divided into blocks, each one 
related to the DM and to the heavy pseudo-Dirac pair, respectively,
as\footnote{The elements related to the pseudo-Dirac pair are labeled by capital letters, such that $\mathcal{P}^{\chi}_{NN^{\prime}}$ is a $2\times2$ matrix.}
\begin{equation}
    \begin{split}
    \mathcal{P}^\chi=&\begin{pmatrix}
        \mathcal{P}^{\chi}_{11} & \mathcal{P}^{\chi}_{1N^{\prime}}\\
        \mathcal{P}^{\chi}_{N1} & \mathcal{P}^{\chi}_{NN^{\prime}}
    \end{pmatrix}\,.
    \end{split}
\end{equation}
The matrices $\mathcal{P}^{R}$ and $\mathcal{P}^L$ are related to different combinations of the self-energy corrections as follows:
\begin{equation}
    \begin{split}
    \mathcal{P}^R\simeq &\left[\Sigma^R_{hh^{\prime}}-\Sigma^R_{hl^{\prime}}\left(\Sigma^R_{ll^{\prime}}\right)^{-1}\Sigma^R_{lh^{\prime}}\right]^{-1}\,,\\
    \mathcal{P}^L\simeq &\Sigma^L_{hh^{\prime}}-\Sigma^L_{hl^{\prime}}\left(\Sigma^L_{ll^{\prime}}\right)^{-1}\Sigma^L_{lh^{\prime}}\,,
    \end{split}
    \label{eq:PX_definitions}
\end{equation}
where the self-energy contributions  have been split into the light (mostly-active) neutrino sector and the heavy (mostly-sterile) one as\footnote{The index $l\,(l^{\prime})$ in Eq.~(\ref{eq:blocks_self}) runs over light neutrino states while $h\,(h^{\prime})$ goes over the DM and heavy pseudo-Dirac neutrinos.}
\begin{equation}
    \Sigma^\chi=\begin{pmatrix}
        \Sigma^\chi_{ll^{\prime}} & \Sigma^\chi_{lh^{\prime}}\\
        \Sigma^\chi_{hl^{\prime}} & \Sigma^\chi_{hh^{\prime}}
    \end{pmatrix}\,.
    \label{eq:blocks_self}
\end{equation}

\titleprl{Dark matter abundance}
Given the DM production rate $\Gamma_{\mathrm{DM}}^h(p,t)$, we can study the evolution of the DM distribution with the following Boltzmann equation~\cite{Lello:2016rvl}:\footnote{This assumes that any other particle species participating in the DM production is in thermal equilibrium in the plasma.}
\begin{equation}
	\frac{d f^h_{\mathrm{DM}}}{dt}=\Gamma^h_{\mathrm{DM}} (p,t) \left[f_{\mathrm{DM}}^{\mathrm{eq}}-f_{\mathrm{DM}}^h\right]\,,
	\label{eq:Boltzmann_Eq}
\end{equation}
where $f^h_{\mathrm{DM}}$ is the DM distribution for 
positive and negative 
helicity and $f_{\mathrm{DM}}^{\mathrm{eq}}$ is the equilibrium Fermi-Dirac distribution. We are interested in the DM production at temperatures $T\gg m_{\mathrm{DM}}$ in which its mass can be neglected, such that $f_{\mathrm{DM}}^{\mathrm{eq}}=\left[e^{p/T}+1\right]^{-1}$. 

It is natural to consider the freeze-in production of DM given that we expect $\Gamma_{\mathrm{DM}}^h(p,t)<H(t)$, with $H(t)$ the Hubble expansion rate. From the theory perspective, the lepton-number conserving limit we are interested in tends to decouple the DM candidate, having very small Yukawa couplings with the SM sector. On the experimental front, the absence of a compelling observation of the DM radiative decay to X-rays sets stringent constraints on the active-DM mixing $\theta_{\alpha\mathrm{DM}}$. Consequently, the DM production rate is expected to be suppressed for $\theta_{\alpha\mathrm{DM}}\ll1$. In this context, we further assume that initially there is no DM and neglect the build-up of its abundance. Therefore, $f_{\mathrm{DM}}^h$ is neglected with respect to the equilibrium distribution in Eq.~(\ref{eq:Boltzmann_Eq}). 

Considering the DM production takes place in the radiation dominated era, it proves useful to study Eq.~(\ref{eq:Boltzmann_Eq}) in terms of the new variables $\tau\equiv M_W/T(t)$ and $y\equiv p(t)/T(t)$, with $M_W$ the $W$-boson mass. The reason is that $y$ does not change under cosmic expansion once the DM distribution has frozen out, which corresponds to $\tau\rightarrow \infty$. Using the relation $dT/dt=-TH(t)$ for an adiabatic expansion, we finally arrive at
\begin{equation}
	\frac{df^h_{\mathrm{DM}}}{d\tau}=\frac{\Gamma_{\mathrm{DM}}^h(y,\tau)}{H(\tau)\tau}f_{\mathrm{DM}}^{\mathrm{eq}}(y)\,.
	\label{eq:Boltzmann_Eq2}
\end{equation}

\begin{figure}
    \includegraphics[width=0.99\linewidth]{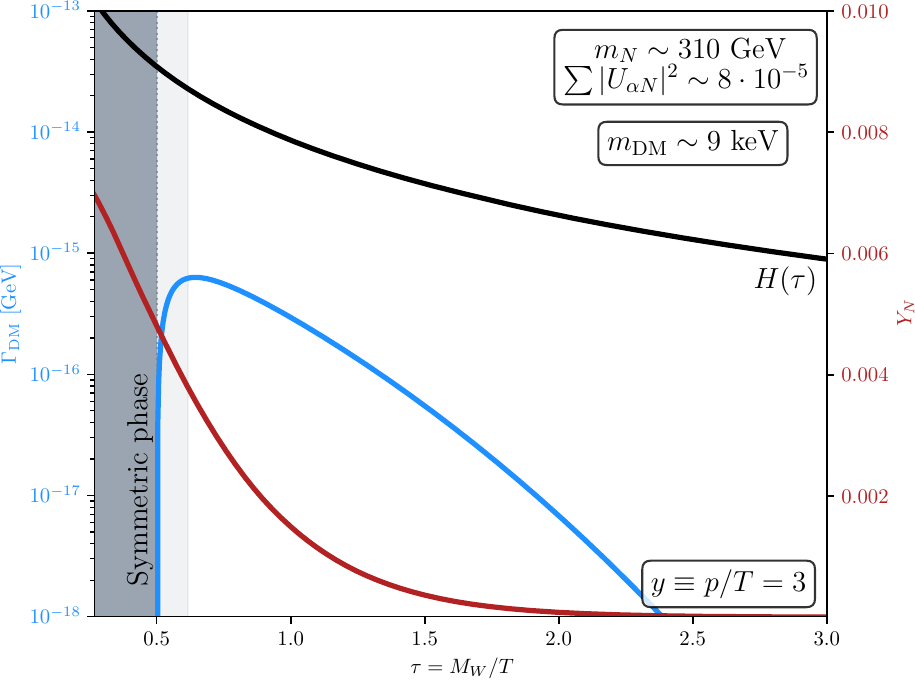}
    \caption{DM production rate (blue line) as a function of $\tau$ for $y=3$. The black line is the Hubble expansion rate, showing $\Gamma_{\mathrm{DM}}\ll H$, while the red one corresponds to the heavy pseudo-Dirac neutrino yield, $Y_N$. The dark gray area highlights the symmetric Higgs phase, while the light gray region corresponds to $\tau<\tau_{\mathrm{in}}$ (see Eq.~(\ref{eq:DM_fraction})).}
    \label{fig:production_Gamma}
\end{figure}

We show in Fig.~\ref{fig:production_Gamma} an example for the DM production rate (in blue), as a function of $\tau$ and for $y=3$, for a set of parameters in the neutrino sector. Clearly, $\Gamma_{\mathrm{DM}}\neq 0$ only after SSB, when neutrino mixing is generated. Moreover, comparing the DM production rate with the Hubble expansion rate (black line), 
we verify $\Gamma_{\mathrm{DM}}\ll H$.\footnote{Values on the left vertical axis should be read for this comparison.} The red line corresponds to the heavy pseudo-Dirac neutrino yield. Its scale is shown in red on the right vertical axis. The dark gray shaded area corresponds to the symmetric phase. Finally, the light gray area corresponds to temperatures very close to the EW crossover, which we choose not to take into account in the analysis.
Once the DM 
has frozen out, we can integrate its distribution over $y$ to find the number density. The fraction of DM produced compared to the observed one~\cite{Planck:2018vyg} is then~\cite{Lello:2015uma}:
\begin{equation}
\begin{split}
	\mathcal{F}_{\mathrm{DM}}=&\left(\frac{\Omega_{\mathrm{DM}}^{\mathrm{obs}}h^2}{0.1198}\right)\frac{m_{\mathrm{DM}}}{7.4\,\mathrm{eV}}\frac{g_{\mathrm{DM}}}{g_d}\times \\
	&\sum_{h=\pm1}\int_0^{\infty} y^2 dy \int_{\tau_{\mathrm{in}}}^{\tau_{\mathrm{fin}}}\frac{df^h_{\mathrm{DM}}}{d\tau}d\tau\,,
\end{split}
\label{eq:DM_fraction}
\end{equation}
where $g_{\mathrm{DM}}$ and $g_d\simeq 100$ are the DM degrees of freedom and the number of relativistic degrees of freedom at decoupling, respectively. In Eq.~(\ref{eq:DM_fraction}), the upper integration limit over $\tau$ corresponds to late enough times such that $\Gamma_{\mathrm{DM}}^h(\tau_\mathrm{fin})\rightarrow 0$. Instead, $\tau_{\mathrm{in}}$ corresponds to the time at which the DM production starts to be effective, which we take $\tau_{\mathrm{in}}\simeq 0.6>\tau_{\mathrm{SSB}}\equiv M_W/T_{\mathrm{SSB}}\simeq 0.5$. 

\titleprl{Existing constraints}
From the neutrino sector, we include the latest NuFIT results~\cite{Esteban:2024eli}, allowing the mass-squared differences to vary in their 95~\% CL ranges. Additionally, the size of the active-heavy mixing can be constrained with different experimental results depending on the heavy neutrino mass scale, $m_N$. For masses below the $Z$-boson mass, there are strong constraints from collider searches in which these heavy neutrinos can be produced~\cite{ATLAS:2019kpx,ATLAS:2022atq,ATLAS:2024fcs,CMS:2018iaf,CMS:2022fut,CMS:2024ake,CMS:2024ita,CMS:2024xdq,Kelly:2021xbv,MicroBooNE:2023eef,Fernandez-Martinez:2023phj}. For larger masses, deviations from unitarity of the leptonic mixing matrix are the leading constraints. We include the bounds obtained at 95~\% CL from a global fit to EWPO and cLFV in Ref.~\cite{Blennow:2023mqx}.

On the other hand, the active-DM mixing can be tightly constrained from indirect detection searches. This DM candidate decays radiatively into a photon and a light neutrino. For the DM masses in the $\mathcal{O}\left(1-100\right)$~keV range that we investigate, its decay produces a monochromatic X-ray signal~\cite{Boyarsky:2007ge,Roach:2019ctw,Foster:2021ngm,Roach:2022lgo}, the non-observation of which constrains $\sin^2\theta_{\alpha\mathrm{DM}}\lesssim 10^{-11}$. Furthermore, light non-cold DM candidates such as sterile neutrinos can alter the primordial power spectrum with respect to the prediction of the standard $\Lambda$CDM model, leaving an imprint on the Ly-$\alpha$ forest~\cite{Gnedin:2001wg,Boyarsky:2008xj}. However, these limits are model dependent as they rely on the DM distribution, which is in turn set by the particular production mechanism. 
Since we consider DM production from the decay of particles in thermal equilibrium, its distribution is related to a thermal one~\cite{Petraki:2007gq,Ballesteros:2020adh}. We take the following lower bound on the DM mass~\cite{Ballesteros:2020adh}:
\begin{equation}
    m_{\rm DM}\geq 7.51\,\mbox{KeV}{\left(\frac{m_{\rm WDM}}{3\,\mbox{KeV}}\right)}^{4/3}{\left(\frac{106.75}{g_d}\right)}^{1/3}\,,
    \label{eq:Lya-forest}
\end{equation}
where $m_{\rm WDM}$ is the bound obtained on the DM mass from Ly-$\alpha$ for a warm dark matter (WDM) candidate~\cite{Narayanan:2000tp,Viel:2005qj,Viel:2013fqw,Baur:2015jsy,Irsic:2017ixq,Palanque-Delabrouille:2019iyz,Garzilli:2019qki}. We will quote the bound using $m_{\mathrm{WDM}}=3$~keV, but we highlight that, given the uncertainties in the rescaling of the WDM limits, weaker constraints can be found using $m_{\mathrm{WDM}}\simeq 1.9$~keV~\cite{Ballesteros:2020adh}.

\titleprl{Numerical analysis}
We have computed $\Gamma_{\mathrm{DM}}^h$ over a lattice in $T$ and $p$, with $T\lesssim 130$~GeV, and integrated Eq.~(\ref{eq:Boltzmann_Eq2}) to obtain the DM abundance today. The DM production rate is only different from zero after SSB, which in our analysis corresponds to $T_{\mathrm{SSB}}\sim 160$~GeV,\footnote{This is in agreement with lattice results~\cite{DOnofrio:2014rug}. More details can be found in the supplemental material.}
as shown in Fig.~\ref{fig:production_Gamma}. We explicitly include both the Higgs vev and mass temperature dependence and
approximate gauge boson masses to their values at $T=0$ given that the Higgs vev becomes large soon after SSB.\footnote{The consistent inclusion of thermal masses for the gauge bosons would need to encompass the dependence of the weak mixing angle with temperature, which is beyond the scope of our work.}

Regarding the parameters in the neutrino sector, we scan over the whole range of the CP-violating phases entering in the PMNS mixing matrix, both Dirac and Majorana. Given that the DM candidate does not substantially contribute to light neutrino masses, the lightest neutrino is massless. The DM and heavy neutrino masses, as well as the complex angles parameterizing $R$ in Eq.~(\ref{eq:Casas-Ibarra}), are varied over the ranges summarized in Table~\ref{tab:ranges_parameters}, where the parameter $\delta\vartheta_{12\,(13)}$ has a different definition depending on the mass ordering: $\delta\vartheta_{12\,(13)}=\vartheta_{12\,(13)}$ for NO or $\delta\vartheta_{12\,(13)}=\vartheta_{12\,(13)}\pm\pi/2$ for IO. 
\begin{table}
\begin{center}
    \begin{tabular}{| c | c | c | c | c |}
    \hline
    $|\delta\vartheta_{12\,(13)}|$ & $\mathrm{Re}\vartheta_{23}$ & $\mathrm{Im}\vartheta_{23}$ & $m_{\mathrm{DM}}$~[keV] & $m_{N}$~[GeV]\\
    \hline
    $\left[10^{-7},\,10^{-2}\right]$ & $\left[10^{-7},\,10^{-2}\right]$ & $[5,\,20]$ & $[5,\,100]$ & $[10,\, 10^5]$ \\
    \hline
    \end{tabular}
    \caption{Parameter ranges for DM and heavy neutrino masses, and the complex angles parameterizing $R$ in Eq.~(\ref{eq:Casas-Ibarra}). In order to have two almost degenerate heavy neutrinos we need $\vartheta_{12\,(13)}= \mp\pi/2+\delta\vartheta_{12\,(13)}$ for IO and $\vartheta_{12\,(13)}= \delta\vartheta_{12\,(13)}$ for NO, while $\mathrm{Im}\vartheta_{23}\gg 1$.}
    \label{tab:ranges_parameters}
\end{center}
\end{table}

After checking the compatibility of a given set of parameters with oscillation data~\cite{Esteban:2024eli} and constraints from X-ray searches~\cite{Boyarsky:2007ge,Roach:2019ctw,Foster:2021ngm,Roach:2022lgo}, we compute the DM abundance to obtain viable regions of parameter space accounting for the observed DM. These are then further constrained by Ly-$\alpha$ forest observations using Eq.~(\ref{eq:Lya-forest}), as well as bounds on the active-heavy mixings~\cite{Blennow:2023mqx} which we parameterize by $\sum|U_{\alpha N}|^2\equiv \sum_{\alpha}\sum_{N=5,\,6}|U_{\alpha N}|^2$. 

\begin{figure}
    \centering
    \includegraphics[width=0.99\linewidth]{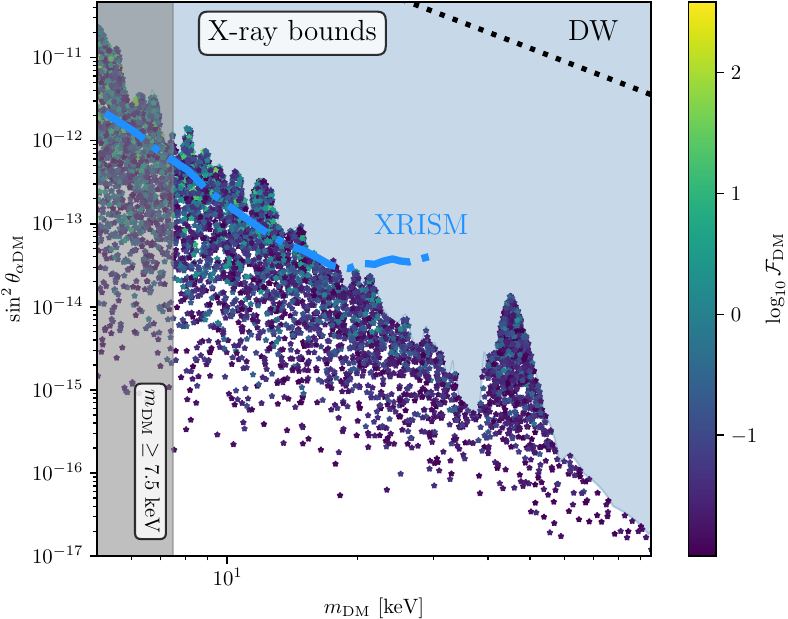}
    \includegraphics[width=0.99\linewidth]{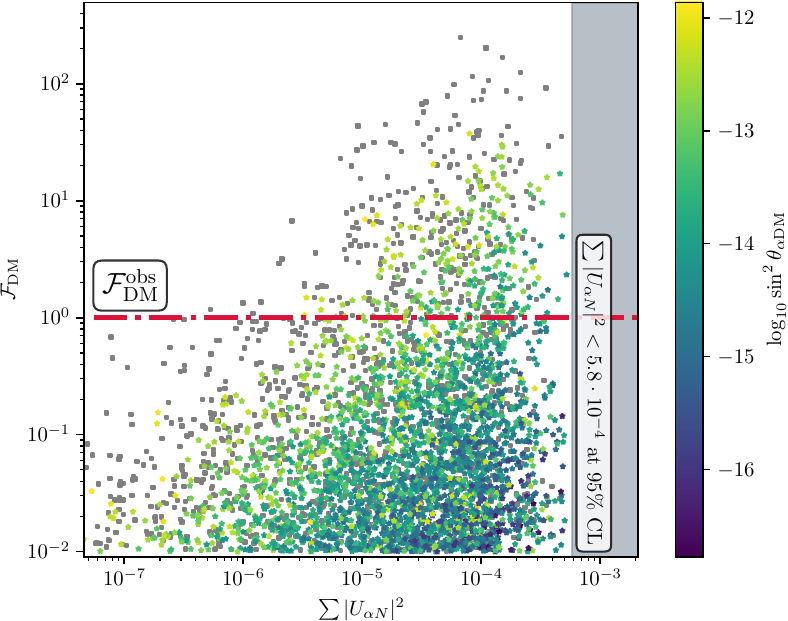}
    \caption{Results for regions of parameter space producing at least $1$~\% of the observed DM for NO. The upper panel shows the results on the DM mixing vs DM mass plane, with the color bar representing the DM fraction. The lower panel shows the DM fraction as a function of the active-heavy mixing. The DM mixing is shown in the corresponding color bar. Relevant experimental and observational bounds are shown in each slice of parameter space (see text for details).}
    \label{fig:DM_abundance_mDM_Mixings}
\end{figure}

We present the results of our scan in Fig.~\ref{fig:DM_abundance_mDM_Mixings}. In the upper panel, we show the DM mixing as a function of its mass. The color code for the points (color bar on the right) 
corresponds to the generated DM fraction. The light blue area corresponds to current X-ray constraints~\cite{Boyarsky:2007ge,Roach:2019ctw,Foster:2021ngm,Roach:2022lgo} while the darker blue dash-dotted line shows the future sensitivity from XRISM~\cite{Dessert:2023vyl}. We show for illustration the region in which the Dodelson-Widrow (DW) mechanism produces all the DM~\cite{Asaka:2006nq} as a dotted black line, already ruled out. The gray shaded area corresponds to the bound on $m_{\mathrm{DM}}$ from Ly-$\alpha$. Noticeably, the total observed DM abundance can only be obtained for $m_{\mathrm{DM}}\lesssim 50\,\mathrm{keV}$. In the lower panel of Fig.~\ref{fig:DM_abundance_mDM_Mixings} we show the dependence of the DM fraction on the size of the active-heavy mixing. In this case the color code represents the size of $\sin^2{\theta_{\alpha\mathrm{DM}}}$. The red horizontal line sets $\mathcal{F}_{\mathrm{DM}}=1$, while gray squared points correspond to regions of parameter space ruled out by Ly-$\alpha$. Although any region above the red line overcloses the Universe and is effectively ruled out, we show it to better understand the dependence on $\sum|U_{\alpha N}|^2$. The gray shaded area corresponds to the bound on $\sum|U_{\alpha N}|^2$ from EWPO and cLFV for NO~\cite{Blennow:2023mqx}. While it is obvious that larger Yukawa couplings translate into a larger DM abundance, we note from this plot that the whole parameter space can be probed with future experiments. On the one hand, constraints pertaining the DM such as X-ray searches or Ly-$\alpha$ tend to close the allowed parameter space over the diagonal.\footnote{Larger values of $\sin^2\theta_{\alpha\mathrm{DM}}$ for a fixed $\sum|U_{\alpha N}|^2$ tend to correspond to larger $\mathcal{F}_{\mathrm{DM}}$.} On the other hand, bounds from EWPO and cLFV on $\sum|U_{\alpha N}|^2$ shut the parameter space in a complementary direction. Similar conclusions are found for the IO case and we do not show the corresponding plots.

\begin{figure}
    \centering
    \includegraphics[width=0.99\linewidth]{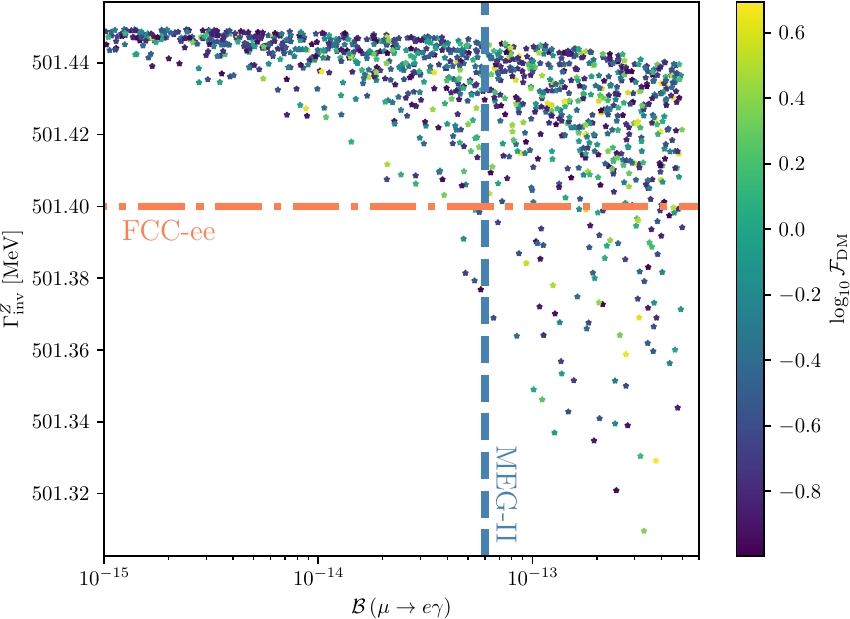}
    \includegraphics[width=0.99\linewidth]{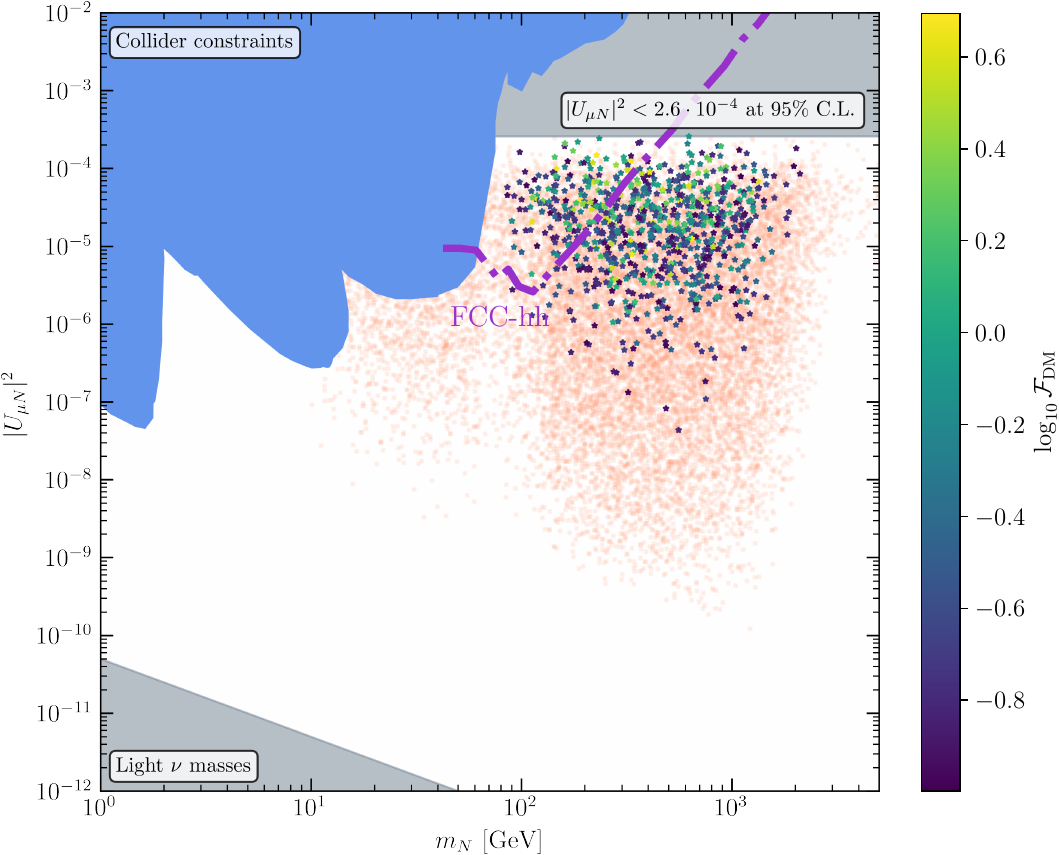}
    \caption{Prospects to probe the parameter space generating $10^{-1}\lesssim \mathcal{F}_{\mathrm{DM}}\lesssim 5$. In the upper panel we show correlations between the invisible decay width of the $Z$-boson and searches for $\mu\rightarrow e\gamma$, together with prospective sensitivities for FCC-ee and MEG-II. In the lower panel we present the relevant parameter space for the active-heavy mixing $|U_{\mu N}|^2$ and $m_N$ including relevant experimental bounds~\cite{Fernandez-Martinez:2023phj,Blennow:2023mqx}. The purple line corresponds to the sensitivity of FCC-hh~\cite{Abdullahi:2022jlv}.}
    \label{fig:results_BR_mutoe-GZinv_Collider}
\end{figure}

Future machines like FCC-ee aim to improve current measurements of EWPO reducing uncertainties by at least one order of magnitude~\cite{FCC:2018evy}, while the quest to find cLFV is still ongoing with the notable example of MEG-II~\cite{MEGII:2018kmf,MEGII:2021fah,MEGII:2023ltw}, searching for $\mu\rightarrow e\gamma$ and currently running. We show in the upper panel of Fig.~\ref{fig:results_BR_mutoe-GZinv_Collider} the consequences large $\sum|U_{\alpha N}|^2$ has on the invisible decay width of the $Z$-boson $\Gamma_{\mathrm{inv}}^Z$, and on $\mathcal{B}\left(\mu\rightarrow e\gamma\right)$, after taking into account existing constraints. The color code (in both panels) represents once again the DM abundance for each point, with $10^{-1}\lesssim \mathcal{F}_{\mathrm{DM}}\lesssim 5$. The orange dash-dotted line represents the potential lower $1\sigma$ region on $\Gamma_{\mathrm{inv}}^Z$ assuming the SM central value and the reduction of current uncertainties by one order of magnitude~\cite{ParticleDataGroup:2024cfk,FCC:2018evy}. Furthermore, we show the prospects from MEG-II~\cite{MEGII:2021fah} with the blue dashed vertical line.
Finally, we show in the lower panel of Fig.~\ref{fig:results_BR_mutoe-GZinv_Collider} our results as a function of the heavy pseudo-Dirac pair mass and their mixing with the muon-neutrino flavor for NO. The shaded blue area corresponds to current collider bounds at $90$~\% CL, obtained using {\tt HNLimits}~\cite{Fernandez-Martinez:2023phj}. The gray horizontal region corresponds to the bounds on $|U_{\mu N}|^2$ from EWPO and cLFV~\cite{Blennow:2023mqx} while the lower gray area corresponds to the naive lower bound on $|U_{\mu N}|^2$ for which the observed mass-squared differences~\cite{Esteban:2020cvm} are generated. The light red cloud of points shows regions of parameter space for which the produced DM abundance is too small ($10^{-5}\lesssim \mathcal{F}_{\mathrm{DM}}\lesssim 10^{-1}$). In order to produce a non-negligible DM abundance we find $100\,\mathrm{GeV}\lesssim m_N\lesssim 1\,\mathrm{TeV}$. Since production is only possible for $T\lesssim 160$~GeV, the DM abundance is exponentially suppressed for $m_N\gtrsim 1\,\mathrm{TeV}$ due to the Boltzmann suppression of the heavy neutrino distribution. Prospects from FCC-hh~\cite{Abdullahi:2022jlv} are shown as a dash-dotted purple line, covering relevant regions of parameter space for $m_N\lesssim 300$~GeV.

\titleprl{Conclusions}
In this letter we proposed a combined solution for the neutrino masses and DM puzzles based on a minimal low-scale seesaw framework, which might also be compatible with leptogenesis. DM production is accounted for through two-body decays of SM bosons, as well as decays of the heavy neutrinos involving DM and a SM boson, at temperatures below the electroweak crossover. For the first time, we perform a complete computation, based on the evaluation of neutrino self-energies in the context of TFT, consistently accounting for all the available production channels, and analyze the phenomenological consequences of such a scenario.

In order for the production to be efficient, approximate lepton number conservation is necessary. This translates into a heavy neutrino spectrum comprised by the DM candidate, with mass $\mathcal{O}\left(10\right)$~keV and almost decoupled, and two heavy Majorana neutrinos with almost degenerate masses, $m_N$, and large mixings with the active ones. We find that the heavy neutrino decay into the DM candidate dominates its production, which translates into the rough upper bound $m_N\lesssim 1$~TeV. Above these masses, the heavy neutrinos would not be abundant enough in the thermal plasma after SSB and the generation of mixings.

The phenomenological implications of such a DM production mechanism are very rich, as it introduces strong synergies between the expected signal in the usual indirect DM probes, such as X-ray searches or constraints from structure formation, and the size of the active-heavy neutrino mixings controlling the final DM abundance. Indeed, current EWPO and searches for cLFV place the leading constraints on this scenario. We find that MEG-II, currently taking data, will be able to probe part of the parameter space for which all the observed DM is generated. In the longer term, the simultaneous improvement of indirect DM searches with experiments like XRISM, as well as the measurement of EWPO in FCC-ee together with searches for cLFV (or even direct searches in FCC-hh) has the potential to completely test this DM generation mechanism.

\titleprl{Acknowledgments}
S.~R.~A sincerely thanks E.~Fernandez-Martinez for insightful discussions on the non-unitarity bounds of the leptonic mixing matrix. He also appreciates J.~Hernandez-Garcia's assistance with {\tt HNLimits} and stimulating exchanges with P.~Hernandez and N.~Rius. Special thanks to Alessandro Granelli for comments on the lepton number conserving limit of the Casas-Ibarra parameterization. M.~L. thanks Fermilab for hosting him during the development of this work. M.~L. is funded by the European Union under the Horizon Europe's Marie Sklodowska-Curie project 101068791 — NuBridge. A.A. acknowledges support from the European Union’s Horizon 2020 research and innovation programme under the Marie Skłodowska -Curie grant agreement No 860881-HIDDeN and the Marie Skłodowska-Curie Staff Exchange  grant agreement No 101086085 – ASYMMETRY.

\bibliography{Biblio}
\bibliographystyle{apsrev4-1}

\newpage
\onecolumngrid
\appendix

\section{Neutrino self-energy corrections}\label{app:lag}
\begin{figure}[h!]
\centering
\begin{tikzpicture}
  \begin{feynman}
    \vertex (a) {$n_i$};
    \vertex [right=1cm of a] (b);
    \vertex [right=3cm of b] (c);
    \vertex [right=1cm of c] (f2) {$n_j$};

    \diagram[horizontal=a to b]{
      (a) -- [plain] (b),
      (b) -- [plain, edge label=$n_k$] (c),
      (b) -- [scalar, edge label=$h$, half left] (c),
      (c) -- [plain] (f2),
    };
  \end{feynman}
\end{tikzpicture}
\begin{tikzpicture}
    \begin{feynman}
    \vertex (a) {$n_i$};
    \vertex [right=1cm of a] (b);
    \vertex [right=3cm of b] (c);
    \vertex [right=1cm of c] (f2) {$n_j$};

    \diagram[horizontal=a to b]{
      (a) -- [plain] (b),
      (b) -- [plain, edge label=$n_k\,(\ell_{\alpha})$] (c), 
      (b) -- [boson, edge label=${Z\,(W)}$, half left] (c),
      (c) -- [plain] (f2),
    };
  \end{feynman}
\end{tikzpicture}
\caption{Feynman diagrams contributing to the neutrino self-energy. In the SM extension with singlet fermions, there is the Higgs contribution (left) and the gauge contribution from the $W$ and $Z$ bosons (right). The imaginary part of the self-energy is related to the rate at which each species approaches equilibrium.}
\label{fig:self-energies}
\end{figure}
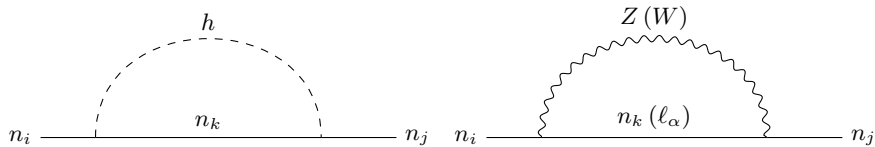

In this section we specify, for completeness, the lagrangian describing the interactions between massive neutrinos and the SM bosons after SSB as well as the diagrams contributing to the neutrino self-energies. We can write the relation between the active neutrinos $\nu_{\alpha L}$, and the mass neutrino eigenstates $n_i$ (with masses $m_i$), as
\begin{equation}
	\begin{split}
		\nu_{L,\,\alpha}=\sum_{i=1}^6 U_{\alpha i} P_L n_i\,
	\end{split}
\end{equation}
where $\alpha=e,\,\mu,\,\tau$ and $i=1,\,2,\,3$ corresponds to light neutrinos while $i=4,\,5,\,6$ to the heavy mostly-sterile ones. Rewriting the interaction lagrangian in the mass basis, taking into account that $n_i$ are Majorana states, we find
\begin{equation}
	\begin{split}
		\mathcal{L}_{W-n}\supset& \frac{g}{\sqrt{2}}W_{\mu}\sum_{i=1}^6\sum_{\alpha}U_{\alpha i}\bar{\ell}_{\alpha}\gamma^{\mu}P_L n_i+h.c.\,,\\
        \mathcal{L}_{Z-n}\supset& \frac{g}{4c_W}Z_{\mu}\sum_{i,j=1}^6 \bar{n}_i \big[C_{ij}\gamma^{\mu}P_L-C_{ij}^*\gamma^{\mu}P_R\big]n_j\,,\\
        \mathcal{L}_{h-n}\supset& -\frac{1}{2v_H}h\sum_{i,j=1}^6 \bar{n}_i \big[C_{ij}\left(m_iP_L+m_jP_R\right)+
		C^*_{ij}\left(m_iP_R+m_jP_L\right)\big]n_j\,,
	\end{split}
	\label{eq:interaction_lagrangian}
\end{equation}
where $\ell_{\alpha}$ are the charged lepton fields, $C_{ij}\equiv \sum_{\alpha} U_{i\alpha}^{\dagger}U_{\alpha j}$, $v_H$ is the Higgs vev, $g$ the $SU(2)_L$ gauge coupling and $c_W$ is a short-hand notation for the cosine of the weak mixing angle. Given these interaction terms, the neutrino self-energy corrections are given by the diagrams in Fig.~\ref{fig:self-energies} involving a Higgs or a gauge boson. Following Ref.~\cite{Weldon:1983jn}, one can relate these self-energy corrections to the different decays we are interested in, namely $n_k\rightarrow h\,(Z)+\mathrm{DM}$ with $m_k\gtrsim M_H\,(M_Z)$, $h\,(Z)\rightarrow n_{k}+\mathrm{DM}$ with $n_k$ a light mostly-active neutrino (or a heavy mostly-sterile one with $m_k<M_H\,(M_Z)$), and $W\rightarrow \ell_{\alpha}+\mathrm{DM}$. 

\section{Derivation of the DM production rate}\label{app:propagating_modes}
The computation of the DM production rate follows an analogous strategy to Refs. \cite{Lello:2016rvl,Abada:2023mib}. We can decompose the neutrino self-energy corrections depicted in Fig.~\ref{fig:self-energies} as
\begin{equation}
    \slashed{\Sigma}^{\chi}=\gamma_0\Sigma^{\chi\,(0)}-\vec{\gamma}\cdot \hat{p}\Sigma^{\chi\,(1)}+\Sigma^{\chi\,(2)}\,,
    \label{eq:self_energy_projection}
\end{equation}
where each $\Sigma^{\chi\,(i)}$ depends on $p^{\mu}$ and $T$. Such self-energy corrections have been determined in~\cite{Abada:2023mib} using the real time formalism of TFT \cite{Schwinger:1960qe,Keldysh:1964ud,le_bellac_1996,kapusta_gale_2006,Lundberg:2020mwu}, to which the reader can refer for more details. Projecting the equation of motion with $P_{L,R}=\left(1\mp \gamma_5\right)/2$ and expanding the fields in helicity eigenstates:
\begin{equation}
    \psi_{\chi}=\sum_{h=\pm 1} v^h \otimes \psi_{\chi}^h,
\end{equation}
with $v^h$ being the eigenstates of the helicity operator with $h=\pm1$ and $\chi\in\lbrace L,\,R\rbrace$, one arrives at the full inverse propagator in the medium:\footnote{We anticipate that the terms proportional to $\Sigma^{\chi\,(2)}$ have a subleading contribution that we neglect, justifying the expression given in the main text.}
\begin{equation}
\begin{split}
    \mathcal{S}^{-1}_h\equiv &p_0^2-p^2-\Omega_h = p_0^2-p^2+\left(p_0-h p\right)\left(\Sigma^{L\,(0)}+h\Sigma^{L\,(1)}\right)\\
    &-\left(p_0-hp\right)\left(\mathcal{M}+\Sigma^{R\,(2)}\right)
    \left[p_0-hp+\Sigma^{R\,(0)}-h\Sigma^{R\,(1)}\right]^{-1}\left(\mathcal{M}+\Sigma^{L\,(2)}\right)\,.
    \label{eq:FullInverse_Propagator}
\end{split}
\end{equation}
Note that this is a matrix equation in ``flavor'' space\footnote{By ``flavor'' here we do not refer explicitly to flavor as $\alpha=e,\,\mu,$ and $\tau$ but rather that there are family indexes involved. It is actually written in the neutrino mass basis, obtained after diagonalizing the mass matrix at $T=0$, such that $\mathcal{M}$ is diagonal.}, such as $p_0$ and $p$ are proportional to the identity matrix, which is not explicitly written. In order to find the propagating states in the medium, one needs to find the simple poles of the propagator, or rather the complex zeroes of its inverse given in Eq.~(\ref{eq:FullInverse_Propagator}). The imaginary part of the complex zeros of $\mathcal{S}_h^{-1}$ for each propagating neutrino species will be related to the rate at which they reach equilibrium~\cite{Weldon:1983jn}. 

The first step to obtain $\Gamma_{\mathrm{DM}}^h$ is to find the eigenvalue of $\Omega_h$ associated to the DM candidate, which we call $\omega_h$, such that we can obtain the dispersion relation of the DM in the medium as $p_0^2-p^2-\omega_{h}=0$. We expect it to behave as $\omega_h\sim m_{\mathrm{DM}}^2+\delta\omega_h$ with $|\delta\omega_h|\ll m_{\mathrm{DM}}^2$. To this purpose, we separate $\Omega_h$ into blocks corresponding to the light mostly-active neutrino species (denoted with indexes $l,\,l^{\prime}=1,\,2,\,3$) and the heavy mostly-sterile ones (denoted by $h,\,h^{\prime}=4,\,5,\,6$):\footnote{We drop the subscript $h$ related to helicity from $\Omega_h$ in Eq.~(\ref{eq:Omegablock}) to ease the notation.}
\begin{equation}
\label{eq:Omegablock}
    \Omega\equiv 
    \begin{pmatrix}
        \Omega_{ll^{\prime}} & \Omega_{lh^{\prime}}\\
        \Omega_{hl^{\prime}} & \Omega_{hh^{\prime}}
    \end{pmatrix}\,.
\end{equation}
We further define $\Sigma^L\equiv \Sigma^{L\,(0)}+h\Sigma^{L\,(1)}$ and $\Sigma^R\equiv \Sigma^{R\,(0)}-h\Sigma^{R\,(1)}$, whose explicit expressions in therms of the neutrino mixings and couplings are
\begin{equation}
\begin{split}
    \left(\Sigma^{R}\right)_{ij}=&\frac{1}{4}\sum_{k=1}^6\left\lbrace \left(\frac{g}{2c_W}\right)^2C_{ik}^*(\sigma^{(0)}_{k,\,Z}-h\sigma^{(1)}_{k,\,Z})C^*_{kj}+\frac{1}{v_H^2}\left(C^*_{ik}m_i+C_{ik} m_k\right)(\sigma^{(0)}_{k,\,H}-h\sigma^{(1)}_{k,H})\left(C_{kj}m_j+C^*_{kj} m_k\right)\right\rbrace \,,\\
    \left(\Sigma^{L}\right)_{ij}=&\frac{g^2}{2} C_{ij}(\sigma^{(0)}_{W}+h\sigma^{(1)}_{W})\\
    +&\frac{1}{4}\sum_{k=1}^{6}\left\lbrace \left(\frac{g}{2c_W}\right)^2C_{ik}(\sigma^{(0)}_{k,\,Z}+h\sigma^{(1)}_{k,\,Z})C_{kj}+\frac{1}{v_H^2}\left(C_{ik}m_i+C^*_{ik} m_k\right)(\sigma^{(0)}_{k,\,H}+h\sigma^{(1)}_{k,H})\left(C^*_{kj}m_j+C_{kj} m_k\right)\right\rbrace\,.
\end{split}
\label{eq:explicit_self_energies}
\end{equation}
In Eq.~(\ref{eq:explicit_self_energies}), $\sigma_{k,\,B}^{(i)}$ is the thermal part of the amplitude for the $k^{\mathrm{th}}$-neutrino mass eigenstate and $B$ corresponds to either the Higgs or Z-boson. For the $W$-boson contribution, $\sigma_W^{(i)}$, we neglect charged lepton masses. The superscript $(i)$ labels the different projections of the self-energy from Eq.~(\ref{eq:self_energy_projection}). Details on the computation of these thermal parts can be found in Ref.~\cite{Abada:2023mib}. Using Eq.~(\ref{eq:blocks_self}) and neglecting light neutrino masses we find that the dominant contributions to $\Omega$ are given by
\begin{equation}
    \Omega=-\left(p_0-hp\right)
    \begin{pmatrix}
        \Sigma^L_{ll^{\prime}} & \Sigma^L_{lh^{\prime}}-\Sigma^{R\,(2)}_{lh^{\prime}}P_{h^{\prime}h}^RM_{\mathrm{heavy}}\\
        \Sigma^L_{hl^{\prime}}-M_{\mathrm{heavy}}P_{h^{\prime}h}^R\Sigma^{L\,(2)}_{hl^{\prime}} & \Sigma^L_{hh^{\prime}}-M_{\mathrm{heavy}}P_{h^{\prime}h}^RM_{\mathrm{heavy}}
    \end{pmatrix}\,,
    \label{eq:explicit_Omega}
\end{equation}
with $P_{h^{\prime}h}^R\equiv \left[p_0-hp+\Sigma^R_{hh^{\prime}}-\Sigma^R_{hl^{\prime}}\left(p_0-hp+\Sigma^R_{ll^{\prime}}\right)^{-1}\Sigma^R_{lh^{\prime}}\right]^{-1}$ and $M_{\mathrm{heavy}}\simeq\mathrm{diag}(m_{\mathrm{DM}},\,m_N,\,m_N)$. At this point, we need to specify the helicity in order to do further simplifications. 
\begin{itemize}
    \item Positive helicity $h=+1$:
    
    We can approximate $p_0-hp\simeq m_{\mathrm{DM}}^2/(2p)$ for the DM candidate, finding as well that $P_{h^{\prime}h}^R\simeq \mathcal{P}^R$ as defined in the main text. Terms proportional to $M_{\mathrm{heavy}}$ dominate and thus we have to solve the following eigenvalue problem
\begin{equation}
    \mathrm{det}\begin{pmatrix}
        -\frac{m_{\mathrm{DM}}^2}{2p}\Sigma^L_{ll^{\prime}}-\omega_{+1} & \frac{m_{\mathrm{DM}}^2}{2p}\Sigma^{R\,(2)}_{lh^{\prime}}\mathcal{P}^RM_{\mathrm{heavy}}\\
        \frac{m_{\mathrm{DM}}^2}{2p}M_{\mathrm{heavy}}\mathcal{P}^R\Sigma^{L\,(2)}_{hl^{\prime}} & \frac{m_{\mathrm{DM}}^2}{2p}M_{\mathrm{heavy}}\mathcal{P}^RM_{\mathrm{heavy}}-\omega_{+1}
    \end{pmatrix}=0\,.
\end{equation}
Using the fact that $\omega_{+1}\sim m_{\mathrm{DM}}^2+\delta\omega_{+1}$, with $|\delta\omega_{+1}|/m_{\mathrm{DM}}^2\ll1$, we can compute the determinant by blocks and focus on the dominant term which is given by
\begin{equation}
    \mathrm{det}\Bigg[\frac{m_{\mathrm{DM}}^2}{2p}M_{\mathrm{heavy}}\mathcal{P}^RM_{\mathrm{heavy}}-\omega_{+1}\Bigg]=0\,,
    \label{eq:full_det_posh}
\end{equation}
where we have dropped terms of order $\mathcal{O}\left(m_{\mathrm{DM}}^2/p^2\right)$. Computing the determinant in Eq.~(\ref{eq:full_det_posh}) by further dividing the heavy sector into the DM and the heavy pseudo-Dirac pair blocks, we arrive at the rate given in Eq.~(\ref{eq:DM_prodRate_hp1}) in the main text by using the relation $\Gamma^{h=+1}_{\mathrm{DM}}=-2\mathrm{Im}\left[\sqrt{p^2+\omega_{+1}}\right]$.

    \item Negative helicity $h=-1$:

    For negative helicity DM we have $p_0-hp\simeq 2p$, and thus the eigenvalue problem we need to solve is instead
\begin{equation}
    \mathrm{det}
    \begin{pmatrix}
        -2p\Sigma^L_{ll^{\prime}}-\omega_{-1} & -2p\Sigma^L_{lh^{\prime}}\\
        -2p\Sigma^L_{hl^{\prime}} & -2p\left(\Sigma^L_{hh^{\prime}}-M_{\mathrm{heavy}}P^{R}_{h^{\prime}h}M_{\mathrm{heavy}}\right)-\omega_{-1}
    \end{pmatrix}=0\,.
\end{equation}
In this case, we can approximate $P^{R}_{h^{\prime}h}\simeq \left(\mathbb{I}_3-\Sigma^R_{h^{\prime}h}/(2p)\right)/(2p)$, such that we need to solve
\begin{equation}
    \mathrm{det}\Big[-2p\left(\Sigma^L_{hh^{\prime}}-M_{\mathrm{heavy}}P^{R}_{h^{\prime}h}M_{\mathrm{heavy}}\right)-\omega_{-1} 
    +4p^2\Sigma^L_{hl^{\prime}}\left[2p\Sigma^L_{ll^{\prime}}+\omega_{-1}\right]^{-1}\Sigma^L_{lh^{\prime}}\Big]=0\,.
\end{equation}
Making use of the fact that $2p\Sigma_{ll^{\prime}}^L\gg m_{\mathrm{DM}}^2$ and neglecting terms suppressed by powers of the momentum, while further dividing between the DM and heavy pseudo-Dirac pair blocks, we arrive at the production rate given in Eq.~(\ref{eq:DM_prodRate_hm1}) of the main text by using the fact that $\Gamma^{h=-1}_{\mathrm{DM}}=-2\mathrm{Im}\left[\sqrt{p^2+\omega_{-1}}\right]$.
\end{itemize}

\section{EW symmetry breaking}\label{app:Thermal_masses}
In order to follow the evolution of EW symmetry breaking as well as the temperature dependence of the Higgs vev and its mass it is necessary to study the Higgs effective potential at finite temperature. Lattice simulations show that the EW crossover starts at $T_{\mathrm{SSB}}\sim160$~GeV and that the Higgs vev approaches its zero temperature value soon after~\cite{DOnofrio:2014rug}. We study the temperature evolution of the Higgs vev and its mass by analyzing the one-loop effective potential including temperature corrections~\cite{Quiros:1999jp}:
\begin{equation}
    V(\langle h\rangle,T)=2D(T^2-T_0^2)\langle h \rangle^2-E T\langle h \rangle^3+\frac{\lambda}{4}\langle h \rangle^4\,.
    \label{eq:effective_potential}
\end{equation}
We explicitly written the potential in Eq.~(\ref{eq:effective_potential}) in terms of the background field $\langle h\rangle$, and the constants $D$, $T_0^2$, and $E$ can be found in Ref.~\cite{Quiros:1999jp} in terms of the physical masses of the gauge bosons and the top quark. In particular, $T_0^2$ sets the temperature at which SSB happens and we set its value to $T_0\sim 160$~GeV, in agreement with lattice results~\cite{DOnofrio:2014rug}. Compared to Ref.~\cite{Quiros:1999jp}, we neglect the temperature corrections in the Higgs quartic coupling given that they largely simplify our analysis while introducing a negligible change.

\begin{figure}
    \centering
    \includegraphics[width=0.49\linewidth]{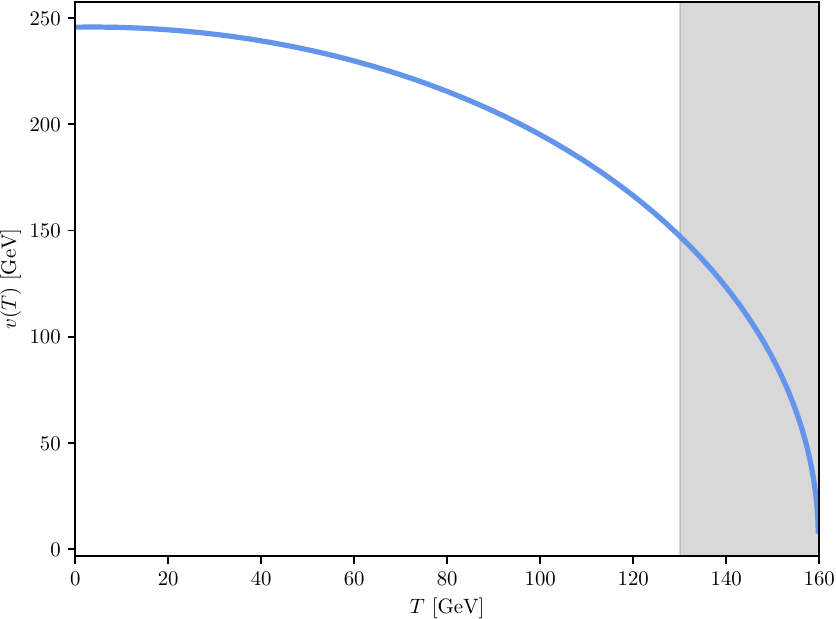}
    \includegraphics[width=0.49\linewidth]{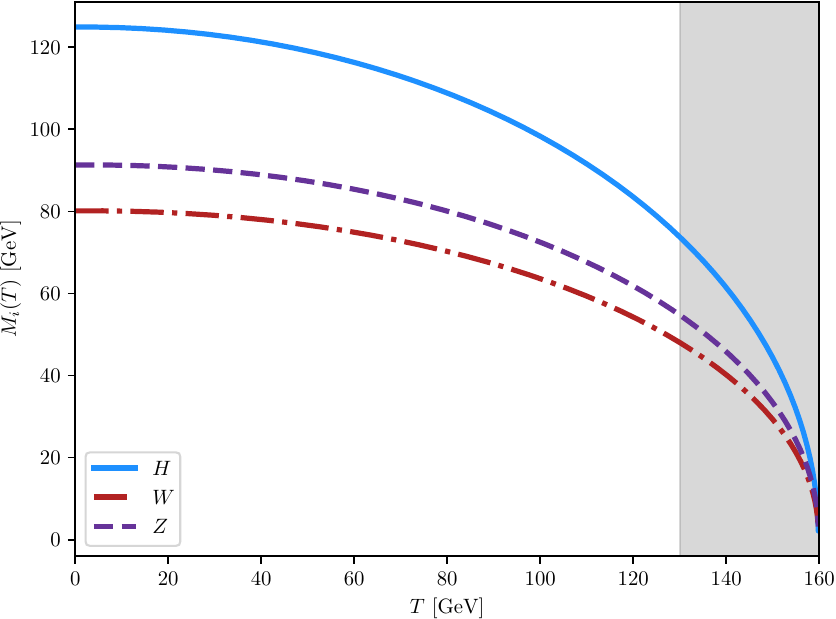}
    \caption{Temperature evolution of the Higgs vev given by Eq.~(\ref{eq:Higgs_vev}) (left panel) and the temperature dependent Higgs mass including the leading thermal corrections (right panel). Additionally, we show the evolution of the gauge boson masses which are proportional to the Higgs vev. The gray shaded area corresponds to temperatures $T\gtrsim 130$~GeV for which we do not compute the DM production.}
    \label{fig:temperature_vev}
\end{figure}

For $T<T_0$, the Higgs develops a vev, $\langle h\rangle=v(T)$, given by
\begin{equation}
    \begin{split}
        v(T)=\frac{3 ET+\sqrt{9E^2T^2-16D\lambda(T^2-T_0^2)}}{2\lambda}\,,
    \end{split}
    \label{eq:Higgs_vev}
\end{equation}
while its mass is found to be
\begin{equation}
    M_H^2(T)\equiv \frac{\partial^2V(\langle h\rangle,T)}{\partial \langle h\rangle^2}\Bigg|_{\langle h\rangle=v(T)}=4D(T^2-T_0^2)-6ETv(T)+3\lambda v(T)^2\,.
    \label{eq:Higgs_mass}
\end{equation}
We show in Fig.~\ref{fig:temperature_vev} the temperature evolution of the Higgs vev (left panel) and its mass (blue line in right panel), which are fully taken into account in our analysis. We also show for illustration the gauge boson mass temperature dependence, which is proportional to the Higgs vev. Note however that the gauge boson thermal masses are not included here. The gray region represents the temperatures above which we do not compute the DM production to avoid a strong dependence on the dynamics of the crossover near the critical temperature.

\end{document}